\documentclass[preprint]{emulateapj}
\usepackage{amstext}
\usepackage{amsmath}
\usepackage{apjfonts}

\shorttitle{GRB050509b}
\shortauthors{Lee, Ramirez--Ruiz \& Granot}

\begin{document}

\title{A compact binary merger model for the short, hard GRB
050509\lowercase{b}}

\author{William H. Lee\footnote{Instituto de Astronom\'{\i}a, UNAM,
Ciudad Universitaria, M\'{e}xico DF 04510}, Enrico
Ramirez-Ruiz\footnote{Chandra Fellow} and Jonathan
Granot\footnote{KIPAC, P.O. Box 20450, Mail Stop 29, Stanford, CA
94309}} \affil{Institute for Advanced Study, Einstein Drive,
Princeton, NJ 08540}

\begin{abstract}
The first X--ray afterglow for a short ($\sim 30\;{\rm ms}$), hard
$\gamma$--ray burst was detected by {\it Swift} on 9 May 2005 (GRB
050509b). No optical or radio counterpart was identified in follow--up
observations. The tentative association of the GRB with a nearby giant
elliptical galaxy at redshift $z=0.2248$ would imply a total energy
release $E_{\gamma, {\rm iso}} \approx 3 \times 10^{48}$~erg, and that
the progenitor had traveled several tens of kpc from its point of
origin, in agreement with expectations linking these events to the
final merger of compact binaries driven by gravitational wave
emission. We model the dynamical merger of such a system and the
time--dependent evolution of the accretion tori thus created. The
resulting energetics, variability, and expected durations are
consistent with GRB 050509b originating from the tidal disruption of a
neutron star by a stellar mass black hole, or of the merger of two
neutron stars followed by prompt gravitational collapse of the massive
remnant.  We discuss how the available $\gamma$-ray and X-ray data
provide a probe for the nature of the relativistic ejecta and the
surrounding medium.
\end{abstract}


\section{Introduction} \label{intro}

Classical $\gamma$-ray bursts (GRBs) naturally divide into two classes
based on their duration and spectral properties \citep{ketal93}:
short/hard ($t<2\;$s), and long/soft ($t>2\;$s) bursts. Through the
impetus of the {\it BeppoSAX} satellite, it became clear that those of
the long variety signal the catastrophic collapse of massive, rapidly
rotating stars \citep{woosley93} at high redshift
\citep{metzger97}. The nature of short events (about 1/3 of the
total), is still undetermined, but the merger of two compact objects
in a tight binary, as will occur in PSR1913+16 \citep{ht75} and
PSRJ0737-3039 \citep{burgay03} in 300~Myr and 85~Myr respectively, has
long been considered a prime candidate for a progenitor
\citep{bp86,elps89}. The short duration of the prompt $\gamma$-ray
emission, however, precluded determination of accurate positions and
follow--up observations, until now.

A breakthrough came on 9 May 2005, when {\it Swift} succeeded in
promptly localizing GRB 050509b, a short burst lasting only $t_{50}
\sim$ 30~ms \citep{gehrels05,Bloom05}. The fast response allowed for
an accurate position determination and a rapidly fading X-ray source
was located, falling below detection within $\sim 300$ s. For the next
few days several multiwavelength observations were made, but
unfortunately no optical or radio afterglow was detected. Although the
issue of a host and its implications for the distance scale remain to
be resolved, initial reports of a giant elliptical galaxy at redshift
$z=0.2248$, lying only $10^{"}$ away from the burst
position\footnote{The projected distance is $\approx 40$~kpc.} appear
consistent with model expectations of compact binary mergers.  This is
because a compact object binary could take hundreds of millions of
years to spiral together, and could by then --- if given a substantial
kick velocity upon formation --- have traveled several tens of
kiloparsecs away from its point of origin \citep[see e.g., ][for
population synthesis estimates]{bsp99,ietal03}. The detection of GRB
050509b thus presents us with the unique opportunity, to which this
{\it Letter} is devoted, to constrain this scenario, both from the
prompt $\gamma$-ray emission and the afterglow. In \S~\ref{energ} we
address the energetics and timescales which can be expected for the
merger of two compact objects based on recent calculations, and
compare them with the data for GRB 050509b.  In \S~\ref{after} we
constrain the properties of the ejecta and the external medium by
using the information available to us from both the afterglow and
prompt emission, considering both the distance scale of the tentative
host galaxy and a higher redshift. Our findings are summarized in
\S~\ref{diss}.

\section{Energetics and Intrinsic Time Scales of the Trigger}
\label{energ} It has long been assumed \citep{ls74} that the merger of a
black hole-neutron star (BH-NS) or double neutron star (NS-NS) binary
would result in the formation of an accretion disk with enough mass
and internal energy to account for the energetics of a typical GRB,
through tidal disruption of the neutron star in the former, or
post--merger collapse of the central core in the latter. Calculations
supporting this view have been carried out in the Newtonian regime,
resulting in disks with $m_{\rm d}\approx 0.3 M_{\odot}$, $kT \approx
10$~MeV, $\rho \approx 10^{11}$g~cm$^{-3}$, which could power a GRB
\citep{rjs96,kl98,rosswogns1}. General Relativity (GR) is certain to
play a role, but gauging its effects is not an easy task. The star
could plunge directly into the black hole and be accreted whole in a
matter of a millisecond \citep{miller05}, precluding the production of
a GRB 10 to 100 times longer.  Pseudo--Newtonian simulations
\citep{rosswog05} and post-Newtonian orbital evolution estimates
\citep{prl04}, however, reveal that the star is frequently distorted
enough by tidal forces that disk--like structures and long, partially
unbound tidal tails can form. The outcome is sensitive to the mass
ratio $q=M_{\rm NS}/M_{\rm BH}$ and it appears that rotating BHs favor
the creation of disks \citep{taniguchi05}. For mass ratios $q\simeq
0.25$ it is possible to form a disk\footnote{For the 18 galactic BH
binaries, an absolute lower bound is $M_{\rm BH} \geq 3.2 M_{\odot}$,
and for 8 of them (44\%), average values yield $6.5 < M_{\rm
BH}/M_{\sun} < 7.5 $ \citep{mr04}.}, although of lower mass than
previously thought, $m_{\rm d}\approx 10^{-2} M_{\odot}$.

To better estimate the mass of the disk (which will crucially affect
the energetics) and the circumstances under which it may form, we have
extended our study of merging BH-NS pairs using a pseudo--Newtonian
potential in three dimensions \citep{lk99b} and summarize our new
results in Table~\ref{BHNS}. A relatively narrow, but not unlikely
range of parameters allows for the formation of a small disk, with
$m_{\rm d} \approx 3 \times 10^{-2} M_{\odot}$. Mass ratios higher
that 1/3 are unlikely to occur, and if $q \leq 0.1$ only a wide,
relatively cold arc--like structure is formed. The densities and
temperatures in the resulting disks are $\rho \approx
10^{10}-10^{11}$g~cm$^{-3}$ and $kT \approx 2-5$~MeV. We have
considered the stiffness of the nuclear equation of state as a
parameter by using polytropes with various indices in the range $5/3
\leq \Gamma \leq 2$. The standard mass for the neutron star is $1.4
M_{\odot}$.

\begin{deluxetable}{cccc}
\tablecaption{Disk formation in BH-NS mergers.\label{BHNS}} 
\tablewidth{0pt} 
\tablehead{\colhead{$M_{\rm NS}/M_{\rm BH}$} 
& \colhead{$\Gamma$} & \colhead{$m_{\rm d}/M_{\sun}$} 
& \colhead{$m_{\rm tail}/M_{\sun}$ } }
\startdata 
0.3 & 5/3 & 0.03 & 0.05 \\
0.2 & 5/3 & 0.03 & 0.05 \\
0.1 & 5/3 & ---  & 0.01 \\
0.3 & 2.0 & 0.04 & 0.1 \\
0.2 & 2.0 & 0.03 & 0.1 \\
0.1 & 2.0 & ---  & 0.02 \\
\enddata
\end{deluxetable}

In the case of merging neutron stars \citep{shibata}, a low--mass disk
(with $\approx$ 1\% of the total mass) may survive once the
supra--massive remnant collapses because of gravitational wave
emission on a time scale shorter than $\approx 100$~ms, and release up
to 10$^{50}$~erg in neutrinos.  In addition, the merger process and
the collapse itself would likely produce a signal of their own
\citep{rosswogns2}.

\begin{figure}
\plotone{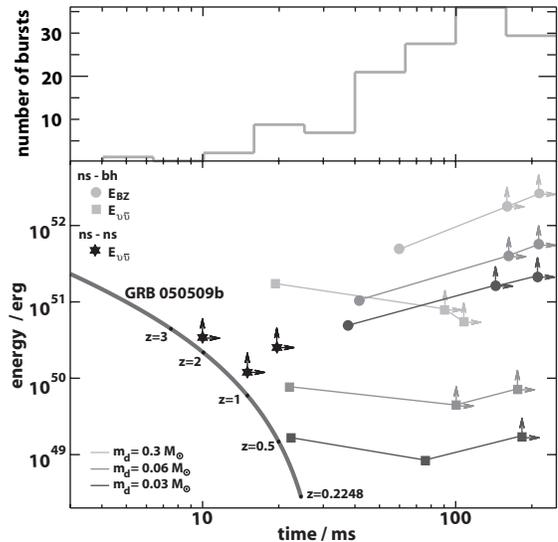}
\caption{{\footnotesize {\it Top:} Histogram of observed short--hard
burst durations taken from \citet{4b}. {\it Bottom}: Comparison of the
energy--duration relation as a function of redshift for GRB 050509b
(black line) with estimates from compact binary mergers. The square
(round) joined symbols show the total isotropic energy release
(assuming collimation of the outflow into $\Omega_{\rm b}=4 \pi /10$)
and duration ($t_{50}$) for $\nu \overline{\nu}$ annihilation
(Blandford--Znajek)--powered bursts, as computed from our 2D disk
evolution models. The range in initial disk mass covers one order of
magnitude and the effective disk viscosities are $\alpha=10^{-1},
10^{-2}, 10^{-3}$ (left to right). Many of the estimates are lower
limits because at the end of our calculations not enough mass had
drained from the disk for the luminosity to drop appreciably. The
stars correspond to $\nu \overline{\nu}$--driven outflows in NS-NS
mergers, computed by \citet{rosswogns3}.}}
\label{fig1}
\end{figure}

Once a disk is formed, the energy output depends on its initial mass,
$m_{\rm d}$ and temperature. We have recently calculated
\citep{lrrp05} a realistic set of time-dependent models for their
dynamical evolution, covering the typical duration time scales of
short GRBs (up to one second). These 2D models make use of a Smooth
Particle Hydrodynamics (SPH) code in azimuthal symmetry and include an
accurate equation of state which considers photodisintegration,
neutronization, and a relativistic Fermi gas of arbitrary degeneracy,
as well as neutrino cooling and finite optical depths to
neutrinos. From the resulting neutrino luminosities we have computed
the total energy deposition that could drive a relativistic outflow
through $\nu \overline{\nu}$ annihilation, assuming a 1\% efficiency
at $L_{\nu}=10^{53}$~erg~s$^{-1}$ \citep{pwf99} and its duration. The
results for various disk masses and effective $\alpha$-disk
viscosities are shown in Figure~\ref{fig1} (joined square symbols),
along with the energy--duration curve for GRB 050509b constrained by
the redshift. The total output $E_{\nu \overline{\nu}} \simeq 10^{49}
[m_{\rm d}/0.03 M_{\sun}]^{2}$~erg, is roughly independent of the
inferred duration, which increases with decreasing disk viscosity
since the overall evolution is slower \citep[see][]{lrrp04}. The
strong dependence on disk mass reflects the sensitivity of the
neutrino emission rates on temperature ($e^{\pm}$ capture on free
nucleons dominates the cooling rate, with emissivity $\dot{q} \propto
\rho T^{6}$).

Magnetically dominated outflows may alternatively tap the disk energy
through the Blandford--Znajek mechanism. Our estimates are shown in
Figure~\ref{fig1} (round symbols), assuming equipartition of the
magnetic field energy density and the internal energy of the fluid in
the inner disk. The energy flux is sensitive primarily to the
equatorial flow density. It is thus initially roughly constant, then
drops on an accretion (i.e. viscous) timescale.  This explains the
energy--duration correlation in Figure~\ref{fig1}.  Since the observed
flux sets the threshold for burst detection, neutrino powered events
will enhance the relative importance of shorter events (since
$E_{\nu\bar{\nu}}\sim{\rm const}$, then $L_{\nu\bar{\nu}}\propto
t^{-1}$), while magnetically dominated short GRBs more truthfully
reflect the underlying intrinsic distribution (since $L_{\rm
BZ}\sim{\rm const}$, then $E_{\nu\bar{\nu}}\propto t$). Relaxing the
assumption of full equipartition will lower the total energy budget
accordingly. The dependence on disk mass is different, with $E_{BZ}
\simeq 5 \times 10^{50} [m_{\rm d}/0.03 M_{\sun}]
[\alpha/10^{-1}]^{-0.55}$~erg. Our estimates assume that whatever seed
field was present has been amplified to the correspondingly high
values extremely rapidly. Whether this will actually occur is unclear,
particularly for the shortest events, as the field can grow only in a
time scale associated with proto--neutron star--like convection or
differential rotation in the case of the MRI.

\section{Constraints on the properties of the ejecta and the 
external medium}
\label{after}

The afterglow of GRB 050509b was detected by the {\it Swift} XRT
during an observation which started $62\;$s after the burst and lasted
$1.6\;$ks (Bloom et al. 2005) with a flux of $F_X\approx 7\times
10^{-13}\;{\rm erg\;cm^{-2}\;s^{-1}}$ in the $0.2-10\;$keV range at
$t=200\;$s, and a temporal decay index $\alpha\approx
1.3^{+0.4}_{-0.3}$, where $F_\nu\propto t^{-\alpha}$.  The numerous
upper limits in the optical and few upper limits in the radio are not
very constraining for the theoretical models (see Bloom et
al. 2005). The fact that the X-ray flux was already decaying at
$t\gtrsim 60\;$s implies $t_{\rm dec}< 60\;$s, where
\begin{equation}\label{t_dec}
t_{\rm dec}=(1+z)\frac{R_{\rm dec}}{2c\Gamma_0^2}=
42(1+z)\left(\frac{E_{51}}{n_0}\right)^{1/3}
\left(\frac{\Gamma_0}{100}\right)^{-8/3}\;{\rm s}
\end{equation}
and $R_{\rm dec}=(3E_{\rm k,iso}/4\pi nm_pc^2\Gamma_0^2)^{1/3}$ are
the observed time and radius where the outflow decelerates
significantly, $\Gamma_0$ is the initial Lorentz factor,
$n=n_0\;{\rm cm^{-3}}$ is the external density and $E_{\rm
k,iso}=10^{51}E_{51}\;$erg is the isotropic equivalent kinetic
energy. That is $\Gamma_0=87[t_{\rm dec}/(1+z)60\,{\rm
s}]^{-3/8}(E_{51}/n_0)^{1/8}$.

\subsection{Prompt Emission from Internal Shocks}
\label{IS}

\begin{figure}
\plotone{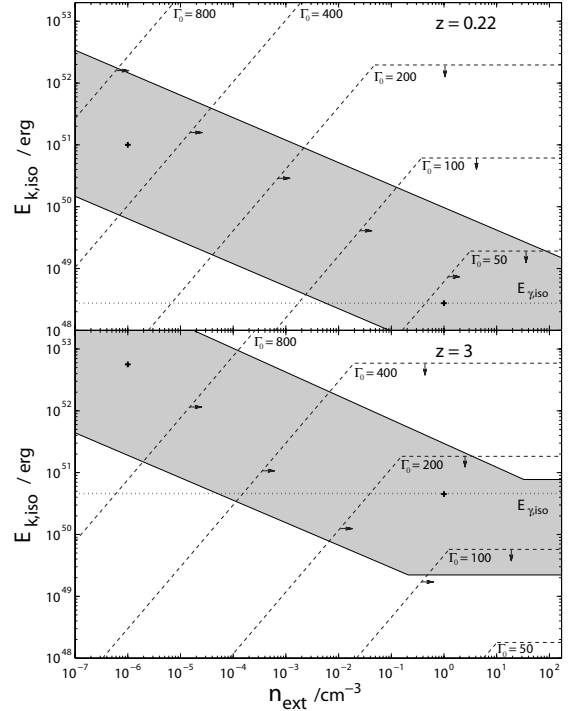}
\caption{{\footnotesize Constraints on the isotropic
equivalent kinetic energy, $E_{\rm k,iso}$, and the external density,
$n$, for the redshift of the tentative host galaxy ($z=0.2248$; {\it
upper panel}) and for $z=3$ ({\it lower panel}). Dashed lines labeled
by the value of the initial Lorentz factor, $\Gamma_0$, bound the
regions of allowed parameter space (in the direction of the
arrows). These limits apply only if the prompt emission is from
internal shocks, and are derived from the requirements that $t_{\rm
dec}< 60\;$s and $\tau_T(R_{\rm IS})<1$. The shaded region is that
allowed from the X-ray flux at the {\it Swift} XRT observation
($t\approx 200\;$s) for a reasonable range of values for the
micro-physical parameters: $2.2 < p <2.5$, $0.03 < \epsilon_E < 0.03$,
$10^{-3} < \epsilon_B < 0.1$ (this is independent of the model for the
prompt emission). The `plus' symbols show the location of the
exemplary models given in Table 3 of Bloom et al.  (2005).}}
\label{fig2}
\end{figure}

Internal shocks typically occur at a radius $R_{\rm IS}\approx
2\Gamma_0^2ct_v$ where $t_v$ is the variability time. Since GRB
050509b had a single peaked light curve \citep{gehrels05}, $t_v\approx
T_{\rm GRB}/(1+z)$ where $T_{\rm GRB}\approx 30\;$ms is the observed
burst duration. The Thompson optical depth is $\tau_T=E_{\rm
k,iso}\sigma_T/4\pi R^2 m_p c^2\Gamma_0$. To see the prompt emission
we need $\tau_T(R_{\rm IS})<1$ implying
\begin{equation}\label{tau_T}
\Gamma_0>100\,E_{51}^{1/5}\left(\frac{t_v}{30\,{\rm
    ms}}\right)^{-2/5}\ .
\end{equation}
For internal shocks, the $\nu F_\nu$ spectrum peaks at
\begin{equation}\label{E_p_IS}
E_{\rm p} = h\nu_m =
\frac{1.3\,g^2}{\sqrt{1+z}}\epsilon_{B,-2}^{1/2}\epsilon_{e,-1}^2
E_{51}^{1/2}\frac{(30\,{\rm ms})^{3/2}}{t_v T_{\rm GRB}^{1/2}}
\left(\frac{\Gamma_0}{100}\right)^{-2}\,{\rm keV}\ ,
\end{equation}
where $\epsilon_B=10^{-2}\epsilon_{B,-2}$ and
$\epsilon_e=0.1\epsilon_{e,-1}$ are the fractions of the internal
energy behind the shock in the magnetic field and in relativistic
electrons, respectively, $g=3(p-2)/(p-1)$ and $p$ is the power-law
index for the electron energy distribution. Eqs. \ref{tau_T} and
\ref{E_p_IS} imply
\begin{equation}\label{E_p_IS2}
E_{\rm p} < \frac{1.3\,g^2}{\sqrt{1+z}}\,\epsilon_{B,-2}^{1/2}\epsilon_{e,-1}^2
E_{51}^{1/10} \left(\frac{t_v}{30\,{\rm ms}}\right)^{-1/5}
\left(\frac{T_{\rm GRB}}{30\,{\rm ms}}\right)^{-1/2}\,{\rm keV}\ .
\end{equation}
The {\it Swift} BAT spectrum is $\nu F_\nu \propto \nu^{0.5\pm 0.4}$
in the $15-350\;$keV range \citep{Barthelmy05}, implying $E_{\rm
p}\gtrsim 300\;$keV, which is hard to achieve for internal shocks (see
Eq. \ref{E_p_IS2}).  Possible ways of increasing $E_{\rm p}$ are if
(i) the internal shocks are highly relativistic, rather than mildly
relativistic as assumed above, or (ii) only a small fraction of the
electrons are accelerated to relativistic energies
\citep[e.g.][]{rrlr02}.  It is not clear how likely either of these
options is.  The constraints on the physical parameters in the
internal shocks model are summarized in Fig. \ref{fig2}.

\subsection{Prompt Emission from the External Shock}
\label{ES}
In this case $t_{\rm dec}\approx T_{\rm GRB}\approx 30\;$ms, implying
a very high $\Gamma_0\approx 1500(E_{51}/n_0)^{1/8}$. Also,
$E_{\rm p}=\max(h\nu_m,h\nu_c)$ where
\begin{eqnarray}\label{nu_m}
h\nu_m &=&
8.8\,g^2\epsilon_{B,-2}^{1/2}\epsilon_{e,-1}^2E_{51}^{1/2}(t_{\rm
  dec}/30\,{\rm ms})^{-3/2}\;{\rm MeV}\ , \\ \label{nu_c} h\nu_c &=&
25(1+Y)^{-2}\epsilon_{B,-2}^{-3/2}n_0^{-1}E_{51}^{-1/2} (t_{\rm
  dec}/30\,{\rm ms})^{-1/2}\;{\rm keV}\ ,
\end{eqnarray}
and $Y$ is the Compton y-parameter. The value of $E_{\rm p}$ is
reasonable and independent of $n$ for $\nu_c<\nu_m$. This requires,
however, sufficiently high values of $n$ and $E_{\rm k,iso}$.

In the external shock model the prompt emission and the afterglow are
produced in the same physical region. It is thus instructive to check
whether the extrapolation of the flux in the prompt emission to the
XRT observation at $t\approx 200\;$s reproduces the observed flux.
The prompt fluence was $f\approx (2.3\pm 0.9)\times 10^{-8}\;{\rm
erg\;cm^{-2}}$ in the $15-350\;$keV BAT range \citep{Barthelmy05}
implying a $\gamma$-ray flux of $F_\gamma(20\,{\rm ms})\approx
10^{-6}\;{\rm erg\;cm^{-2}\;s^{-1}}$. The spectral slope of $\nu
F_\nu\propto\nu^{0.5\pm 0.4}$ implies an X-ray flux of $F_X(20\,{\rm
ms})\approx 2\times 10^{-7}\;{\rm erg\;cm^{-2}\;s^{-1}}$ in the
$0.3-10\;$keV XRT range. This, in turn, implies an average temporal
decay index of $\langle\alpha\rangle\approx 1.3-1.4$ between $20\;$ms
and $200\;$s. One might expect $\langle\alpha\rangle$ to be somewhat
smaller, as the maximal value of $\alpha$ is $(3p-2)/4$ (i.e. $1.375$
for $p=2.5$)\footnote{This is valid before the jet break time
\citep{GS02}, which most likely occurs significantly later than
$200\;$s.} at $\nu>\max(\nu_m,\nu_c)$ which is above $300\;$keV at
$20\;$ms. This results in overproducing the flux at $200\;$s by a
factor of $\sim 10-20$ for $p=2.5$. The observed flux is reproduced
for $p\approx 2.8$. Lower values of $p$ might still be possible if,
e.g., there are significant radiative losses or a much higher
$\epsilon_B$ in the very early afterglow.

\section{Discussion and Prospects}
\label{diss}

From the inferred energy per solid angle, simple blast-wave models
seem able to accommodate the data on the afterglow of GRB 050509b.
Constraints on the angle-integrated $\gamma$-ray energy are not very
stringent --- the outflow could be concentrated in a high Lorentz
factor beam only a few degrees across, or actually be wider.  Standard
arguments concerning the opacity of a relativistically expanding
fireball \citep{bp86} indicate that Lorentz factors $\Gamma \gtrsim
10^{2}$ are required, with a baryon loading no larger than $\sim
10^{-4} M_{\sun}$.  As we have argued in \S~\ref{IS}, for GRB 050509b
internal shocks face the problem of explaining the observed peak
energy. With an external shock, the required Lorentz factor is high by
usual standards, and such accelerations would accordingly require a
remarkably low baryon loading close to the central engine.

Only detailed simulations in full GR will provide us with the details
of the merger process in a compact binary. However, an approximate
treatment using variable compressibility in the equation of state and
a range of mass ratios leads to similar outcomes, suggesting that the
creation of a dense torus is a robust result. If the central engine
involves such a configuration, is it possible to discriminate between
the alternate modes for its formation: compact merger or collapsar?
Accurate localizations of further events should help to confirm or
reject the latter option, since a collapsar would occur in or near a
region of recent star formation, contrary to the expectations
concerning compact object mergers (see \S~\ref{intro}). A more direct
test would obviously be a detection, or lack thereof, of a
supernova--like signature\footnote{It is important to note that the
natural time scale for a collapsing envelope to produce a GRB is given
by the fall-back time, which is longer than a few seconds.}
\citep{Bloom05,Hjorth05}. Definitive and spectacular confirmation
could come from the detection of a coincident gravitational wave
signal in the 0.1--1 kHz range, since mass determinations in X--ray
binaries and the binary pulsars indicate that in NS-NS systems mass
ratios should be close to unity, whereas in BH-NS binaries they should
be smaller than 1/3. Accurate measurement of the inspiral waveform in
the LIGO band would allow simultaneous determination of the ratio of
reduced to total system mass, $\mu/M_{\rm T}$ and of the "chirp" mass
$M_{\rm c}=(m_1 m_2)^{3/5}/(M_{\rm T})^{1/5}$, from which the mass
ratio can be derived. 

GRB 050509b is the first event in the short class of bursts for which
we have an accurate localization and a tentative distance indicator,
based on an association with an elliptical galaxy at $z=0.2248$. At
the inferred distance of $\simeq 1$~Gpc, we have shown here that the
energetics and duration can be accounted for by small, dense disks
around stellar mass black holes, based on dynamical modeling of such
systems. The lack of a SN--like signature in the optical at that
distance \citep{Hjorth05} argues against a collapsar/hypernova
progenitor. Putting GRB 050509b at a significantly higher redshift
places more serious constraints not only due to the energetics, but
particularly because of the short duration: at $z=3$, $E_{\gamma, {\rm
iso}} \approx 4.6 \times10^{50}$~erg and $t_{50} \approx 8$~ms (see
Figures~\ref{fig1} and \ref{fig2}). This is hard to reconcile with the
current models, and makes it unlikely that a collapsing stellar core
is at the origin of GRB 050509b. The observed duration distribution of
bursts may be affected by the mechanism responsible for the production
of the relativistic outflow, with magnetically powered events more
faithfully reflecting the intrinsic population. GRB 050509b is in many
respects an unusual event, being so short and apparently
sub--energetic.

Much progress has been made in understanding how $\gamma$-rays arise
from the sudden deposition of energy in a small volume, and in
deriving the properties of the afterglows that follow. The identity of
short--burst progenitors remains a standing mystery, which further
observations of events similar to GRB 050509b will hopefully help
elucidate.

\acknowledgments We thank J. Bloom, J. Hjorth, C. Kouveliotou,
P. Kumar, D. Page, D. Pooley, J. Prochaska and S. Rosswog for helpful
conversations.  This work is supported by CONACyT-36632E (WHL), the
DoE under contract DE-AC03-76SF00515 (JG), and NASA through a Chandra
Fellowship award PF3-40028 (ER-R).

\end{document}